\RequirePackage{fix-cm}
\documentclass[smallextended]{svjour3}       % onecolumn (second format)
\usepackage{amsmath}
\smartqed  % flush right qed marks, e.g. at end of proof
\usepackage{graphicx}
\usepackage{multirow}
\usepackage{epsfig}
\usepackage{epstopdf}
\usepackage{amsfonts}

\journalname{}

\title{The Golay codes and Quantum Contextuality}

\author{Mordecai Waegell$^{1}$ and P.K. Aravind$^{2}$}

\authorrunning{M.Waegell, P.K. Aravind}

\institute{M.Waegell$^{1}$,P.K.Aravind$^{2}$ \at
$^{1}$Institute for Quantum Studies, Chapman University, Orange, CA.
$^{2}$Physics Department, Worcester Polytechnic Institute, Worcester, MA 01609, U.S.A.\\
\email{waegell@chapman.edu,paravind@wpi.edu}}

\date{\today}

\begin{document}
\maketitle
\begin{abstract}

It is shown that the codewords of the binary and ternary Golay codes can be converted into rays in $\mathbb{RP}^{23}$ and $\mathbb{RP}^{11}$ that provide proofs of the Kochen-Specker theorem in real state spaces of dimension 24 and 12, respectively. Some implications of these results are discussed.

\end{abstract}
\section{\label{sec:Intro}Introduction}

In 1949 Golay\cite{Golay},\cite{Conway},\cite{Sloane},\cite{Preskill} discovered two remarkable error correcting codes, a binary code, now designated by the symbol\footnote{The symbol we use for classical error-correcting codes is standard, and consists of three numbers placed within square brackets. We will use bold font for this symbol whenever it occurs, so that it is not confused with the references.} \textbf{[24,12,8]}, consisting of $2^{12}=4096$ codewords of 24 characters (each a 0 or 1) with a minimum distance\footnote{The (Hamming) distance between two codewords is the number of places in which they differ.} of 8 between the codewords, and a ternary code, with symbol \textbf{[12,6,6]}, consisting of $3^{6}=729$ codewords of 12 characters (each a 0,1 or 2) with a minimum distance of 6 between the codewords.\footnote{These codes are actually ``extended'' codes obtained from the codes \textbf{[23,12,7]} and \textbf{[11,6,5]} of length 23 and 11, respectively, by the addition of a parity check digit. The latter codes are sometimes referred to as ``punctured'' codes.} These codes led to significant advances in cryptography and mathematics in the decades following their discovery. In cryptography the Golay codes have the distinction of being the only perfect codes over finite fields that can correct more than a single error in their codewords\footnote{The maximum number of errors that can be corrected in either of the binary codes is three, and in the ternary codes two.}.  In mathematics the binary Golay code led to the discovery of the remarkable Leech lattice\cite{Leech} in 24 dimensions that provides the densest packing of identical spheres in this dimension\cite{Cohn} (the only other dimension in which such a packing is known being 8). And in group theory, as Preskill\cite{Preskill} put it, the Golay codes set in motion the entire sequence of events that led to the complete classification of the finite groups (including particularly the ``sporadic'' groups) in the last part of the previous century.\\

The advent of quantum computing and the resulting interest in quantum error correction led to a renewed interest in classical cryptography when it was realized that many of the results of the latter could be adapted and put to use in the quantum context. When the first quantum error correcting codes were proposed, it was natural to wonder if in addition to their practical utility they could be used to shed light on any of the fundamental mysteries of quantum mechanics. This question seems to have been first considered by DiVincenzo and Peres\cite{DiV}, who answered it in the affirmative\footnote{It is interesting to note that Shor\cite{Shor} went in the reverse direction by using the three qubit GHZ-Mermin codewords, which had been used to prove the Bell and Kochen-Specker theorems, to propose one of the earliest quantum error correcting codes.}. They showed that the 5-qubit codewords representing logical bits in the quantum code proposed in \cite{Bennett} and \cite{LaFlamme} could be used to give proofs of the Bell nonlocality\cite{Bell} and Kochen-Specker-Bell\cite{KS} theorems, in the spirit of the proofs that had been given earlier by Greenberger, Horne and Zeilinger\cite{GHZ} and Mermin\cite{Mermin}. They pointed out that the 7-qubit codewords of Steane\cite{Steane} also gave rise to such proofs and surmised that other quantum codewords would as well, since they invariably involve entangled states of three or more qubits. Because of the last observation, the connection between quantum error correcting codes and quantum paradoxes (i.e. the Bell and Kochen-Specker theorems), though of great interest, is perhaps not entirely unexpected.\\

But is there any connection between classical error correcting codes and the quantum paradoxes mentioned above? No obvious example comes to mind. It is the purpose of this paper to show that the codewords of the two Golay codes can be converted into rays in state spaces of dimension 24 and 12 that provide proofs of the Kochen-Specker (KS) theorem in these dimensions. The proofs are simple, and follow from the impossibility of solving certain Diophantine equations.\\

Since numerous proofs of the KS theorem in all dimensions from 3 up are known\cite{Budroni}, the discovery of yet another example might not occasion much surprise. Nevertheless the present demonstration may be of interest because it reveals a surprising connection between classical error correcting codes and quantum contextuality, two subjects that are not usually thought of as being related to each other. A further discussion of this point will be given after the new results have first been presented.\\

\section{\label{sec:binary}The binary Golay code and the KS theorem in 24 dimensions}

Figure \ref{tab1} shows a generator matrix for the binary Golay code. The $2^{12} = 4096$ codewords can be obtained from it as

\begin{equation}\label{Eq1}
 w(n) = \sum_{i=1}^{12} a_{n,i} v_i  \hspace{1mm}  \hspace{3mm} ,
\end{equation}

\noindent
where $v_{i}$ is the i-th row of the matrix, $a_{n,i} \in (0,1)$, the sum over $i$ is done bitwise modulo 2 and we have introduced the integer 

\begin{equation}\label{Eq2}
  n = \sum_{i=1}^{12} a_{n,i}2^{12-i}+1 \hspace{1mm} 
\end{equation}

\noindent
to label the codeword $a_{n,1}a_{n,2}\cdots a_{n,12}$ (note that $n$ is one more than the decimal integer defined by the codeword).\\ 

We next convert each codeword into a ray in $\mathbb{RP}^{23}$ by replacing each 1 in it by a $-1$ and each 0 by a 1. Since the codewords $n$ and $4097-n$ are the complements of each other (i.e., have their 0's and 1's exchanged), they map into the same ray, which we will denote by the integer $n$. The codewords of the binary Golay code then give rise to a system of 2048 rays in $\mathbb{RP}^{23}$, which we will label by the integers 1 through 2048. These rays form a large number of bases (i.e.,  sets of 24 mutually orthogonal rays), and it is these bases that we must look at to find proofs of the KS theorem.\\

\begin{table}[ht]
\centering % used for centering table
\begin{tabular}{|c |c|} % centered columns (5 columns)
\hline % inserts single horizontal line
 1 0 0 0 0 0 0 0 0 0 0 0 & 1 0 1 0 0 0 1 1 1 0 1 1   \\
 0 1 0 0 0 0 0 0 0 0 0 0 & 1 1 0 1 0 0 0 1 1 1 0 1   \\
 0 0 1 0 0 0 0 0 0 0 0 0 & 0 1 1 0 1 0 0 0 1 1 1 1   \\
 0 0 0 1 0 0 0 0 0 0 0 0 & 1 0 1 1 0 1 0 0 0 1 1 1   \\
 0 0 0 0 1 0 0 0 0 0 0 0 & 1 1 0 1 1 0 1 0 0 0 1 1   \\
 0 0 0 0 0 1 0 0 0 0 0 0 & 1 1 1 0 1 1 0 1 0 0 0 1   \\
 0 0 0 0 0 0 1 0 0 0 0 0 & 0 1 1 1 0 1 1 0 1 0 0 1   \\
 0 0 0 0 0 0 0 1 0 0 0 0 & 0 0 1 1 1 0 1 1 0 1 0 1  \\
 0 0 0 0 0 0 0 0 1 0 0 0 & 0 0 0 1 1 1 0 1 1 0 1 1  \\
 0 0 0 0 0 0 0 0 0 1 0 0 & 1 0 0 0 1 1 1 0 1 1 0 1  \\
 0 0 0 0 0 0 0 0 0 0 1 0 & 0 1 0 0 0 1 1 1 0 1 1 1 \\
 0 0 0 0 0 0 0 0 0 0 0 1 & 1 1 1 1 1 1 1 1 1 1 1 0  \\
\hline
\end{tabular}
\caption{Generator matrix for the binary Golay code (taken from Ref. \cite{Conway}), with the 12 x 12 identity matrix split off at the left. The rows are numbered 1 to 12 from top to bottom and the codewords can be constructed from them via Eq.(\ref{Eq1}).}
\label{tab1} % is used to refer this Figure in the text
\end{table}

A proof of the KS theorem\cite{Peres,Budroni} requires finding a set of bases such that each of the rays in them cannot be assigned the value 0 or 1 in a noncontextual\footnote{The term ``noncontextual'' means that each ray assumes the same value in all the bases in which it occurs.} manner in such a way that every basis has exactly one ray assigned the value 1 in it\footnote{The sets of bases yielding proofs of the KS theorem are often simply referred to as KS sets. We will use the terms KS proofs and KS sets interchangeably in this paper.}. However the number of bases formed by the 2048 rays is so large (in the tens of thousands, at the very least) that any hope of finding a subset of them that yields a proof of the KS theorem seems futile. But we have found a way around this obstacle.\\

First note two elementary facts: (i) two codewords at a distance of 12 from each other correspond to orthogonal rays, and (ii) a set of 24 codewords, any two of which are a distance 12 apart, make up a basis. \\

Now each codeword has 2576 others at a distance of 12 from it, and it is not hard to pick out a basis from this set.  One such basis is shown in the first row of Figure 2, with each ray represented by its integer label. From this ``seed'' basis, one can generate a system of 2048 bases by adding the ray $n$, for $1 \le n \le 2048$, to each ray of the basis (to do this, replace each ray by its codeword, add the codewords bitwise modulo 2 and replace the resulting codeword by its ray.)  It is obvious that this procedure generates bases, because adding the same codeword to two codewords does not change the distance between them. If one lists all the bases obtained in this way starting from the seed basis, one gets a table like that in Figure \ref{tab2}, in which only a few of the rows (or bases) are shown. This table can be thought of as a $2048 \times 24$ matrix, whose elements are rays and whose rows are bases. \\

\small
\begin{table}[ht]
\begin{gather*}
1 \hspace{1mm} 127 \hspace{1mm} 128 \hspace{1mm} 136 \hspace{1mm} 177 \hspace{1mm} 414 \hspace{1mm} 586 \hspace{1mm} 788 \hspace{1mm} 866 \hspace{1mm} 911 \hspace{1mm} 1005 \hspace{1mm} 1011 \hspace{1mm} \hspace{1mm} 1225 \hspace{1mm} 1323 \hspace{1mm} 1324 \hspace{1mm} 1366 \hspace{1mm} 1491 \hspace{1mm} 1510 \hspace{1mm} 1589\hspace{1mm} 1607 \hspace{1mm} 1704 \hspace{1mm} 1722 \hspace{1mm} 1756 \hspace{1mm} 1821 \hspace{1mm} \\
2 \hspace{1mm} 128 \hspace{1mm} 127 \hspace{1mm} 135 \hspace{1mm} 178 \hspace{1mm} 413 \hspace{1mm} 585 \hspace{1mm} 787 \hspace{1mm} 865 \hspace{1mm} 912 \hspace{1mm} 1006 \hspace{1mm} 1012 \hspace{1mm} \hspace{1mm} 1226 \hspace{1mm} 1324 \hspace{1mm} 1323 \hspace{1mm} 1365 \hspace{1mm} 1492 \hspace{1mm} 1509 \hspace{1mm} 1590\hspace{1mm} 1608 \hspace{1mm} 1703 \hspace{1mm} 1721 \hspace{1mm} 1755 \hspace{1mm} 1822 \hspace{1mm} \\
3 \hspace{1mm} 125 \hspace{1mm} 126 \hspace{1mm} 134 \hspace{1mm} 179 \hspace{1mm} 416 \hspace{1mm} 588 \hspace{1mm} 786 \hspace{1mm} 868 \hspace{1mm} 909 \hspace{1mm} 1007 \hspace{1mm} 1009 \hspace{1mm} \hspace{1mm} 1227 \hspace{1mm} 1321 \hspace{1mm} 1322 \hspace{1mm} 1368 \hspace{1mm} 1489 \hspace{1mm} 1512 \hspace{1mm} 1591\hspace{1mm} 1605 \hspace{1mm} 1702 \hspace{1mm} 1724 \hspace{1mm} 1754 \hspace{1mm} 1823 \hspace{1mm} \\
               \cdots          \hspace{6mm}          \cdots   \hspace{6mm}          \cdots   \\
2048 \hspace{1mm} 1922 \hspace{1mm} 1921 \hspace{1mm} 1913 \hspace{1mm}1872\hspace{1mm} 1635\hspace{1mm} 1463\hspace{1mm} 1261\hspace{1mm} 1183\hspace{1mm} 1138\hspace{1mm} 1044\hspace{1mm} 1038\hspace{1mm}  824 \hspace{1mm} 726\hspace{1mm}  725\hspace{1mm}  683\hspace{1mm}  558\hspace{1mm}  539\hspace{1mm}  460\hspace{1mm}  442\hspace{1mm}  345\hspace{1mm}  327\hspace{1mm}  293\hspace{1mm}  228\hspace{1mm}
\end{gather*}
\caption{The 2048 bases obtained from a ``seed'' basis (the one in the first row) by adding all the rays to it in the manner described in the text.}
\label{tab2} % is used to refer this table in the text
\end{table}

\normalsize
It follows from the fact that the Golay code is a linear code that each ray occurs exactly once in every column of the matrix, and therefore 24 times over the entire matrix. Thus this system of rays and bases can be described by the symbol $2048_{24} - 2048_{24}$, in which the numbers represent the rays and bases and the subscripts the incidence of each type of object with the other.\\

But the proof of the KS theorem follows immediately from this symbol because the total number of bases, 2048, is not divisible by the number of bases in which each ray occurs, or 24, making it impossible to have just one ray with the value 1 in each basis.\\   

\noindent
Some comments are in order about this proof :\\

\noindent
(a) It may not be the most economical proof, in that there may be a subset of the bases that yields a briefer proof. Exploring all the subsets is an onerous task, both in view of the large number of bases and the high dimensionality of the space involved, so we proceeded as follows. We picked $n$ mutually orthogonal rays and looked at all rays orthogonal to them to get a subset of the 2048 rays that live in a $(24-n)$-dimensional space, and then tried to find a KS proof in the resulting system of rays and bases. For $n \geq 4$ the number of rays and bases shrink to the point where this possibility is easily ruled out, but for $n \leq 3$ (i.e., in dimensions 21 through 23) the number of bases is large enough that the matter cannot be settled without a more detailed investigation.\\

\noindent
(b) The proof we gave above is based on the bases of Table 2. However the rays of the Golay code give rise to many other bases, and it is possible that the complete set of bases may house both a larger number and greater variety of KS proofs than the particular set we considered. It would particularly interesting to find the smallest proof (in terms of the number of contexts) yielded by this code. \\

\noindent
(c) The ``punctured'' code \textbf{[23,12,7]}, obtained from the binary code by dropping the parity check digit, is of no interest in connection with the KS theorem because if its codewords are converted into rays in $\mathbb{RP}^{22}$ in the same manner as before, no orthogonal pairs of rays result. \\

\section{\label{sec:binary}The ternary Golay code and the KS theorem in 12 dimensions}

Figure \ref{tab3} shows a generator matrix for the ternary Golay code. The $3^{6} = 729$ codewords can be obtained from it as

\begin{equation}\label{Eq3}
  \sum_{i=1}^{6} a_{n,i} v_i  \hspace{3mm}  ,
\end{equation}

\noindent
where $v_{i}$ is the i-th row of the matrix, $a_{n,i} \in (-1,0,1)$ and the addition is done tritwise modulo 3. The label $n$ refers to the codeword $a_{n,1}\cdots a_{n,6}$, but we do not introduce an integer for it as we have no need for it below.\\

\begin{table}[ht]
\centering % used for centering table
\begin{tabular}{|c |c|} % centered columns (5 columns)
\hline % inserts single horizontal line
 1 0 0 0 0 0  & 0 1 1 1 1 1    \\
 0 1 0 0 0 0  & $\bar{1}$ 0 1 $\bar{1}$ $\bar{1}$ 1 \\
 0 0 1 0 0 0  & $\bar{1}$ 1 0 1 $\bar{1}$ $\bar{1}$ \\
 0 0 0 1 0 0  & $\bar{1}$ $\bar{1}$ 1 0 1 $\bar{1}$  \\
 0 0 0 0 1 0  & $\bar{1}$ $\bar{1}$ $\bar{1}$ 1 0 1  \\
 0 0 0 0 0 1  & $\bar{1}$ 1 $\bar{1}$ $\bar{1}$ 1 0  \\
\hline
\end{tabular}
\caption{Generator matrix for the ternary Golay code (taken from Ref.\cite{Conway}), with the 6 x 6 identity matrix split off at the left. The rows are numbered 1 to 6 from top to bottom and the codewords can be constructed from them via Eq.(\ref{Eq3}). Note: $\bar{1}=-1$.}
\label{tab3} % is used to refer this table in the text
\end{table}

If one ignores the codeword with all 0's, the others come in pairs that are the negatives of each other, and keeping only one member of each pair gives a system of 364 rays in $\mathbb{RP}^{11}$. A computer program shows that these rays form 140647 bases (or sets of 12 mutually orthogonal rays), with 132 rays occurring in 9496 bases, 220 rays in 27 bases and 12 rays in 35696 bases, so that the system can be characterized by the symbol $132_{9496} 220_{27} 12_{35696} - 140647_{12}$ (note that the sum of the products of the numbers and their subscripts on the left equals the similar product on the right). But this symbol gives a proof of the KS theorem for the following reason: \\

\noindent
A noncontextual hidden variables theory requires a 1 to be assigned to a certain number of rays of each of the three types so that every basis has exactly one ray assigned the value 1 in it. But if $x,y$ and $z$ are the numbers of rays of the three types assigned a 1, a successful value assignment requires that there be at least one solution to the Diophantine equation $9496x + 27y + 35696z = 140647$ subject to the constraints $0\leq x \leq 132, 0\leq y \leq 220$ and $0\leq z \leq 12$. However it is easily checked that no such solution exists, and this proves the theorem.\\

It is actually possible to give a much quicker proof of the theorem by considering only the 220 rays arising from the 440 codewords of weight 9, which form a total of 495 bases, with each ray occuring in exactly 27 bases (see the Supplementary Material \cite{Supp} for a listing of these rays and their bases). Thus this system can be described by the symbol $220_{27} - 495_{12}$. However since the total number of bases, 495, is not divisible by the number of bases in which each ray occurs, or 27, the theorem is proved. \\ 

It would be interesting to find the smallest KS proof provided by the ternary Golay code. It may occur within the 495 bases formed by the codewords of weight 9 or the much larger set of bases formed by all the codewords. The ``punctured'' code \textbf{[11,6,5]} leads to no bases in 11 dimensions, but it does lead to a large number of bases in 10 dimensions or less and so may harbor KS sets there. We also examined an alternative mapping of the words of the ternary Golay code into rays, in which the digits 0,1 and -1 are mapped into the cube roots of unity, but found that it fails to lead to any proofs of the KS theorem\cite{Megill}.\\

\section{\label{sec:binary} Discussion}

We have shown that the codewords of the binary and ternary Golay codes can be mapped into rays in real Hilbert spaces of dimension 24 and 12 that provide proofs of the Kochen-Specker theorem in these dimensions. The proofs work by showing that the solutions to certain Diophantine equations, supplemented by a constraint in one case, do not exist. Table \ref{tab4} gives an overview of the three proofs presented in the paper.\\

\begin{table}[ht]
\centering % used for centering table
\begin{tabular}{|c|c|c|} % centered columns (5 columns)
\hline % inserts single horizontal line
 Code  & Rays-Bases & Diophantine equation   \\
 \hline
                 &     &                         \\
 Binary Golay code \textbf{[24,12,8]}  & $2048_{24}-2048_{24}$ & $24x = 2048$ \\
  &All rays and the 2048 bases obtained by& \\
                    & adding all the codewords to a fixed basis&                         \\
 \hline
                 &     &                         \\
 Ternary Golay code \textbf{[12,6,6]}  & $132_{9496} 220_{27} 12_{35696} - 140647_{12}$ & $9496x + 27y + 35696z = 140647$ \\
    &  All rays and their bases  &    \\
& & $0\leq x \leq 132, 0\leq y \leq 220$, $0\leq z \leq 12$\\
\hline
                &     &                         \\
Ternary Golay code \textbf{[12,6,6]}  & $220_{27}-495_{12}$ & $27x = 495$ \\
    & Rays corresponding to codewords  & \\
     &of weight 9 and their bases &                         \\
\hline
\end{tabular}
\caption{KS proofs based on the binary and ternary Golay codes. The first column lists the code from which the proof is derived, the second the system of rays and bases used in the proof and the third the Diophantine equation, together with any constraints, whose insolubility constitutes the proof.}
\label{tab4} % is used to refer this table in the text
\end{table}

Despite the fact that it is over half a century old, the Kochen-Specker theorem continues to be of great interest because of the issue of contextuality it has brought to the fore. A recent article by Budroni et. al.\cite{Budroni} provides a comprehensive overview of the KS theorem and contextuality from a contemporary viewpoint, discussing both how the notion of contextuality has evolved over the years and the many uses to which it has been put. On the foundational front, early work on contextuality focused mainly on finding definitive proofs of it based on contradictions between the answers to certain sets of yes-no questions, whereas current efforts have shifted more in the direction of demonstrating it unambiguously in the face of theoretical objections\cite{Meyer} to the definitive proofs as well as the practical limitations of experimental tests. This shift of focus has led to the development of noncontextuality inequalities\cite{KCBS} as well as operationally motivated definitions of noncontextuality\cite{Spekkens} that can lead to more decisive experimental tests\cite{Exp}. As far as its applications are concerned, contextuality has been shown to be the source of speedup in quantum computations of various kinds\cite{speedup}, to guarantee the security of quantum key distribution  protocols in a device independent manner\cite{crypto}, and to be a resource for random number generation\cite{random}, among other things\cite{Budroni}.\\

\begin{figure}
\includegraphics[width=.80\textwidth]{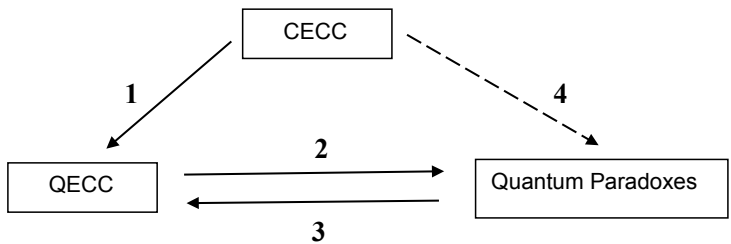}
\caption{CECC = Classical error correcting codes, QECC = Quantum error correcting codes and Quantum Paradoxes = Bell and KS theorems. The connection indicated by  arrow 1 has been made in many ways, notably by Steane\cite{Steane} in his 7-qubit code and later in the CSS codes\cite{CSS}. The connection 3 was made by Shor\cite{Shor}, whose showed how the GHZ-Mermin proof could be made to yield quantum codewords, while several examples of the reverse connection 2 were pointed out by DiVincenzo and Peres\cite{DiV}. The connection 4 has been made in this paper, with the dashed arrow being used because we are aware of only two instances of it (based on the binary and ternary Golay codes).}
\label{tab5}
\end{figure}

Against this backdrop, what is the interest and significance of the new results presented here? They clearly belong to the category of definitive proofs mentioned above, in which interest has now waned because so many of them are known. Nevertheless we would like to argue that our results are of interest for at least a couple of reasons. For one thing, definitive proofs still offer one route to the formulation of noncontextuality inequalities and so new proofs are worth noting in case they lead to better experimental tests or are more suited to a particular application. However a more significant reason, in our opinion, is that they reveal a surprising connection between classical error correcting codes and the seemingly unrelated subject of quantum contextuality. This connection, which is indicated by the dashed arrow in Figure \ref{tab5}, completes the triangular linkage between the three concepts shown there.\\

One might wonder if there are other classical codes that allow the connection indicated by the dashed arrow to be made. A clue to this is provided by the following generalization of the construction we gave earlier based on the binary Golay code. Consider a binary linear code with the symbol \textbf{[2n,k,d]}. This code can be made to yield $2^{k-1}$ rays in $\mathbb{RP}^{2n-1}$ by the same method that was used for the Golay code. Now, if the code has $2n$ codewords that are at a distance of $n$ from each other, then the rays corresponding to these words form a basis and successively adding each of the $2^{k-1}$ rays to this basis\footnote{Recall that the addition must be done by converting the rays into codewords and adding them bitwise modulo 2.} yields a total of $2^{k-1}$ bases in which each of the rays occurs exactly $2n$ times. But if $2^{k-1}$ is not divisible by $2n$, no satisfactory $0/1$ assignment to the rays is possible and this proves the KS theorem. To summarize, the KS theorem can be proved by finding a binary linear code, \textbf{[2n,k,d]}, for which $2^{k-1}$ is not divisible by $2n$ and which also has $2n$ codewords at a distance of $n$ from each other.\\

One possible candidate of this kind is the binary quadratic residue code \textbf{[48,24,12]}, whose generator matrix is given in \cite{Esmali}. This code has many words that differ from each other in exactly half their digits, and thus form pairs of orthogonal rays, but we have not been able to pick out a complete basis from them. If only this step could be supplied,\footnote{Our problem is that we don't know any way, aside from tedious computation, of picking out a basis from the generator matrix of a linear code. We would certainly welcome any help or advice on this point.} we would have a KS proof. It would be nice to settle this issue one way or another and also identify other binary codes that might provide such proofs. The extension of this construction to ternary and n-ary codes is far from obvious. \\

Preskill\cite{Preskill} has pointed out that the binary Golay code can be converted into a quantum error correcting code via the CSS construction. Paralleling this, Prakash\cite{Prakash} recently demonstrated that the ternary Golay code can be converted into a 11-qutrit quantum error correcting code that is useful for magic state distillation and fault tolerant quantum computing. These applications of the Golay codes to current problems in quantum computing may lend additional interest to the results of this paper, which show how the Golay codes can be used, with only a slight modification of their original form, to address a challenge that arose much earlier in the history of quantum mechanics.\\   

If one notes that $12 = 2^{2}3$, $24 = 2^{3}3$ and $48 = 2^{4}3$, these dimensions are the natural successors to $3 = 2^{0}3$ and $6 = 2^{1}3$. Dimension 3 is interesting because it is the lowest dimension in which the KS theorem holds and dimension 6 has the distinction of being the one in which the most compact KS proof (in terms of the number of contexts) is known\cite{Lisonek2014a}. Lison{\u e}k\cite{Lisonek2014b} used a generalization of complex Hadamard matrices to show how the construction in \cite{Lisonek2014a} could be extended to many even dimensions, including the cases of 12 and 24 studied in this paper. However his approach leads to KS sets in the complex spaces $\mathbb{CP}^{23}$ and $\mathbb{CP}^{11}$ that are not directly related to the KS sets found here, although they have the potential to be much smaller. It would be interesting to carry out a comparative study of the KS sets in the two approaches and see, in particular, which can be pushed to yield the smallest sets in dimensions 12 and 24. \\

An important feature of the Golay codes is their large symmetry group, which are Mathieu groups\cite{Conway},\cite{Sloane},\cite{Preskill}. This symmetry is largely inherited by the system of rays and bases in which we have looked for KS sets. Our past experience with highly symmetrical systems (in dimensions of the form $2^{n}$) has shown that they possess a large number of KS proofs with a rich and varied taxonomy\footnote{KS proofs derived from the 600-cell\cite{600} and Gosset's eight-dimensional polytope\cite{E8} have just these features.}. It would be interesting to see if these features persist in the proofs obtained from the Golay codes.\\

Our proofs based on the Golay codes emerged after we first looked for such proofs in the Leech lattice, but gave up. To understand why the Leech lattice might be a good place to look, recall that the only dimensions in which the optimum dense packing of spheres is known are 8 and 24. In 8 dimensions the optimal packing is provided by the E8 lattice, and the 240 vectors from any lattice point to its nearest neighbors yield a system of 120 rays in $\mathbb{RP}^{7}$ that give rise to an astronomical number of KS proofs\cite{E8},\cite{Lisonek2014c}. This led us to expect, by analogy, that the 196,560 vectors from a point of the Leech lattice to its nearest neighbors would give rise to a similarly large number of KS proofs. An investigation showed that this system possesses an extremely large number of bases, but the sheer size of the problem and the absence of any intuition of how to go about looking for KS proofs  made us abandon it. However we feel that a properly designed attack on the Leech lattice, perhaps based on graph theoretical techniques like those in \cite{Ramanathan} or \cite{Severini}, could reveal interesting information about the contextuality buried in it.\\  

Finally, since the sporadic groups grew out of the Golay code and the Leech lattice, it is tempting to speculate that they might also be a fertile breeding ground for quantum contextuality in the high dimensional spaces in which they operate.\\

%\bibliographystyle{ieeetr}
%\bibliography{testbib}

\end{document}